\documentclass[aps,prc,preprint,groupedaddress]{revtex4-1}
\usepackage{graphicx}
\usepackage{epsfig}
\usepackage{geometry}
\usepackage{fancyhdr,amsmath,amssymb,hyperref}
\usepackage{longtable}
\begin{document}

%\preprint{}

\title{The $^{136}$Xe + $^{208}$Pb reaction: A test of models of multi-nucleon transfer reactions}


\author{J. S. Barrett, R. Yanez, W. Loveland}
%\thanks{}
\affiliation{Department of Chemistry, Oregon State University, Corvallis, Oregon 97331 USA}
\author{S. Zhu,  A. D. Ayangeakaa, M. P.  Carpenter, J. P. Greene, R. V. F. Janssens, T. Lauritsen }
\affiliation{Physics Division, Argonne National Laboratory, Argonne, Illinois 60439 USA}
\author{E. A. McCutchan, A. A.  Sonzogni }
\affiliation{National Nuclear Data Center, Brookhaven National Laboratory, Upton, New York, 11973 USA}
\author{C. J. Chiara}
\altaffiliation{Current Address: U.S. Army Research Laboratory, Adelphi, Maryland 20783}
\affiliation{Physics Division, Argonne National Laboratory, Argonne, Illinois 60439 USA}
\affiliation{Department of Chemistry and Biochemistry, University of Maryland, College Park, Maryland  20742 USA}
\author{J. L.  Harker}
\affiliation{Physics Division, Argonne National Laboratory, Argonne, Illinois 60439 USA}
\affiliation{Department of Chemistry and Biochemistry, University of Maryland, College Park, Maryland  20742 USA}
\author{ W. B.  Walters}
\affiliation{Department of Chemistry and Biochemistry, University of Maryland, College Park, Maryland  20742 USA}

\date{\today}

\begin{abstract}
The yields of over 200 projectile-like fragments (PLFs) and target-like fragments (TLFs) from the interaction of (E$_{c.m.}$=450 MeV) $^{136}$Xe with a thick target of  $^{208}$Pb were measured using Gammasphere and off-line $\gamma$-ray spectroscopy,  giving  a comprehensive picture of the production cross sections in this reaction.The measured yields were compared to predictions of the GRAZING model and the predictions  of Zagrebaev and Greiner using a quantitative metric, the theory evaluation factor, {\bf tef}.  The GRAZING model predictions are adequate for describing the yields of nuclei near the target or projectile but grossly underestimate the yields of all other products.  The predictions of Zagrebaev and Greiner correctly describe the magnitude and maxima of the observed TLF transfer cross sections for a wide range of transfers ($\Delta$Z = -8 to $\Delta$Z = +2).  However for $\Delta$Z =+4, the observed position of the maximum in the distribution is four neutrons richer than the predicted maximum.  The predicted yields of the neutron-rich  N=126 nuclei exceed the measured values by two orders  of magnitude.  Correlations between TLF and PLF yields are discussed.

\end{abstract}

%\keywords{}

\maketitle

\section{Introduction}

There has been a renewed interest in the use of multi-nucleon transfer reactions to produce heavy neutron-rich (n-rich) nuclei, motivated by a series of calculations by Zagrebaev and Greiner \cite{z1,z2}.  These reactions were extensively studied experimentally in the 1980s \cite{30,31,32,33,34,35}. (An in-depth review of these data is found in \cite{jens}.) One observed the production of n-rich, trans-target nuclides up to Fm and Md with cross sections $\sim$0.1 $\mu$b.  The basic problem in making heavier nuclei was that the higher excitation energies that led to broader isotopic distributions caused the highly excited nuclei to fission.  The contribution of Zagrebaev and Greiner is to emphasize the role of shell effects in these transfer reactions.  For example, in the reaction of $^{238}$U with $^{248}$Cm, at a modest energy above the Coulomb barrier (1.1 V$_{B}$), Zagrebaev and Greiner predict  a net particle transfer from $^{238}$U to $^{248}$Cm, forming $^{208}$Pb from $^{238}$U and adding 30 nucleons to $^{248}$Cm.  This calculation, when applied to the reaction of E$_{c.m.}$=750 MeV $^{238}$U + $^{248}$Cm, reproduced the previous measurements of Sch{\"a}del et al. \cite{33} and predicted the formation at picobarn levels of new n-rich isotopes of Sg. 

However, these reactions are difficult to study due to the low cross sections and the low intensities of the heavy beams.  But Zagrebaev and Greiner, to their credit,  have provided suggestions of a number of surrogate reactions involving larger cross sections and projectiles with much higher beam intensities  that can be used to test their predictions.  This paper deals with one of those surrogate reactions, the reaction of $^{136}$Xe with $^{208}$Pb at E$_{c.m.}$ = 450 MeV.

\subsection{$^{208}$Pb region}
The study  of multi-nucleon transfer (MNT) reactions to produce nuclei near $^{208}$Pb, i.e., near the N=126 shell closure and an {\it r}-process waiting point, is of interest as a testing ground for models of multi-nucleon reactions and because of the character of the product nuclei.  There is considerable interest in the nuclear spectroscopy community  in making the nuclei ``south" of $^{208}$Pb .  (One motivation is to study the ``quenching" of the shell gap by increasing neutron excess).  The predicted \cite{z1} overall pattern of nuclidic yields in the $^{136}$Xe + $^{208}$Pb reaction (E$_{c.m.}$ = 450 MeV) is shown in Figure 1. The formation of several n-rich nuclei with significant cross sections is predicted.  The formation of several unknown nuclei, such as  $^{202}$Os, is also predicted \cite{z1}.  Further studies \cite {z3} have indicated that multi-nucleon transfer reactions such as $^{136}$Xe + $^{208}$Pb have larger cross sections for producing neutron-rich heavy nuclei than fragmentation of relativistic heavy ions. However, while recent experiments \cite{beli} have shown the nuclide production cross sections for multi-nucleon transfer reactions exceed those of fragmentation reactions, the overall production rates are higher in the fragmentation reactions due to higher beam intensities, target thicknesses and collection efficiencies.   The overall kinematics and mass distributions for the $^{136}$Xe + $^{208}$Pb reaction have been measured \cite{kozulin} and agree quite nicely with the Zagrebaev and Greiner predictions.  Also, contrary to most expectations, Zagrebaev and Greiner show that it is the head-on collisions rather than the grazing collisions that contribute to the yields of the heavy neutron-rich nuclei.  Thus, separators that collect recoils at small angles may be useful in studying MNT reactions \cite{heinz}.

Other laboratories have found enhanced formation (relative to the Zagrebaev and Greiner predictions) of trans-target n-rich nuclei in the reaction of $^{64}$Ni with $^{207}$Pb \cite{heinz} and $^{136}$Xe + $^{208}$Pb \cite{kozulin}.   However, in each of these experiments, the enhanced cross sections were associated with a small number of reaction products.  No large region of enhanced cross sections was found despite predictions of the existence of such regions.  Given the large (and expensive) effort required to pursue the study of these transfer reactions with heavy nuclei and the limited success in this effort  in the 1980Õs, a more compelling case for further studies is needed.   

\subsection {GRAZING}

Another calculational model for multi-nucleon transfer reactions is the semi-classical model of Winther \cite{w1,w2} expressed in the computer code GRAZING \cite{nanni}.  In this model, one uses classical trajectories of the colliding ions (grazing trajectories) and then uses quantal methods to study the probability of exciting collective states in the colliding nuclei and the probability of nucleon transfer.  Multi-nucleon transfer takes place via a multi-step exchange of single nucleons via stripping and pickup.  The code has known shortcomings, i.e., when used to predict the yields of heavy fragments, it does not take into account decay by fission.  (Recently Yanez and Loveland \cite{rw} have developed a modification of GRAZING called GRAZING-F which takes into account fission decay of the primary fragments, removing this deficiency.) Also, in the initial nucleus-nucleus interaction, deformation of the nuclei is not considered.  Nonetheless, the main features of collisions such as $^{64}$Ni + $^{238}$U are adequately described \cite{cor}.  In this work, we shall use the version of GRAZING implemented on the Nuclear Reactions Video Project website \cite{nrvp} with the standard input parameters given on that site.  
\subsection{This experiment}

Large gamma-ray arrays have been used previously to determine the production cross sections of trans-target nuclei in multi-nucleon transfer reactions \cite{krol1,krol2}.  In these experiments, thick targets were irradiated with projectiles that stopped in the targets.  In-beam $\gamma$-$\gamma$ coincidence analysis was used to determine the yields of stable nuclei and beam-off coincidence analysis (between beam pulses) was used to determine the yields of short-lived nuclei.  Standard gamma-ray spectroscopy was then applied to study the decay of the long-lived radioactive products.  The  number of radionuclide yields that can be measured in such experiments is very large and low cross sections (10 $\mu$b) can be measured.  Also, it should be noted that over 130 n-rich nuclei were populated in the $^{64}$Ni + $^{208}$Pb reaction studied with Gammasphere using the thick target technique \cite{wilson}.  It is this technique that we have used to characterize the yields of projectile-like and target-like fragments including  trans-target and neutron-rich nuclei in the $^{136}$Xe + $^{208}$Pb reaction. This technique will only allow the identification of known nuclei, so the use of multi-nucleon transfer reactions to produce new nuclei is not tested.

{\bf We report on the use of Gammasphere to determine the yields of the known target-like and projectile-like fragments from the interaction of $^{136}$Xe with $^{208}$Pb at E$_{c.m.}$ = 450 MeV using the gamma-ray spectroscopy protocols discussed above. We compare the measured fragment yields with the predictions of Zagrebaev and Greiner and the GRAZING code to test these models of multi-nucleon transfer.}

\section{Experimental}
This experiment took place at the Gammasphere facility of the Argonne National Laboratory.  A beam of 785 MeV $^{136}$Xe struck a 49 mg/cm$^{2}$ $^{208}$Pb target (99$\%$ enriched) mounted at the center of Gammasphere.  The beam was stopped in the thick target and the center of target beam energy was 743 MeV (E$_{c.m.}$= 450 MeV).  The ``effective" target thickness (the portion of the target where the beam energy went from the entrance energy of 785 MeV to the reaction barrier energy (700.5 MeV)) was 3.113 mg/cm$^{2}$.  Simulations using GRAZING show the center of target energy product distribution is the best representation of the weighted product distributions in the target.  The intensity of the beam striking the target was monitored periodically by inserting a suppressed Faraday cup into the beam line in front of the target.  The length of the irradiation was 92.1 hours with an average beam current of 0.47 pnA.  

A fully instrumented Gammasphere has 100 Compton-suppressed Ge detectors.  For this run, there were 90 operational Ge detectors.  The measurement strategy was similar to that of \cite{krol1}.  With the beam on, the spacing between the accelerator beam bursts was 824 ns.  Triple $\gamma$-ray coincidence events ($\gamma$-$\gamma$-$\gamma$) were recorded.  After the irradiation was stopped, Gammasphere was switched to singles mode and the target was counted for 39 hours.  Then the target was removed from Gammasphere and $\gamma$-ray spectroscopy of the target radioactivities was carried out using a well-calibrated Ge detector in the ATLAS hot chemistry laboratory.  The total observation period was 5.3 days, during which 15 measurements of target radioactivity were made.  The analysis of these Ge $\gamma$-ray decay spectra was carried out using the FitzPeaks \cite{jim} software.  The end of bombardment (EOB) activities of the nuclides were used to calculate absolute production cross sections, taking into account the variable beam intensities using standard equations for the growth and decay of radionuclides during irradiation \cite {FKMM}.  These measured absolute nuclidic production cross sections are tabulated in the Appendix.  These cross sections represent ``cumulative" yields, i.e., they have not been corrected for the effects of precursor beta decay. These cross sections are identified as being from radioactive decay (RD) in Tables 1 and 2 in the Appendix.  The yields of ground state and isomeric states for a given nuclide were summed to give a total nuclidic cross section.  These cumulative yields are the primary measured quantity in this experiment.  

To correct for precursor beta decay, we have assumed the beta-decay corrected independent yield cross sections for a given species, $\sigma$(Z,A), can be represented as a histogram that lies along a Gaussian curve.
\begin{equation}
\sigma (Z,A)=\sigma (A)\left[ 2\pi C_{Z}^{2}(A)\right] ^{-1/2}\exp\left[ \frac{-(Z-Z_{mp})^{2}}{2C_{Z}^{2}(A)}\right] 
\end{equation}
where $\sigma$(A) is the total isobaric yield (the mass yield), C$_{Z}$(A) is the Gaussian width parameter for mass number A and Z$_{mp}$(A) is the most probable atomic number for that A.  Given this assumption, the beta decay feeding correction factors for cumulative yield isobars can be calculated once the centroid and width of the Gaussian function are known.  

To uniquely specify $\sigma$(A), C$_{Z}$(A), and Z$_{mp}$(A), one would need to measure three independent yield cross sections for each isobar.  That does not happen often.  Instead one assumes the value of  $\sigma$(A) varies smoothly and slowly as a function of mass number and is roughly constant within any A range when determining C$_{Z}$(A), and Z$_{mp}$(A). The measured nuclidic formation cross sections are then placed in  groups according to mass number.    We assume the charge distributions of neighboring isobaric chains are similar and radionuclide yields from a limited mass region can be used to determine a single charge distribution curve for that mass region.  One can then use the laws of radioactive decay to iteratively correct the measured cumulative formation cross sections for precursor decay.  These ``independent yield" cross sections are also tabulated in Tables 1 and 2.  The cumulative and independent yield cross sections are similar due to the fact that without an external separation of the reaction products by Z or A, one most likely detects only a single or a few nuclides for a given isobaric chain and these nuclides are located near the maximum of the Gaussian yield distribution.  As a byproduct of this procedure, one also derives the mass distribution for the reaction $\sigma$(A). which are the independent yield cross sections divided by the fractional chain yields.  These mass yields are shown in Figure 4.

The analysis of the post-beam decay measurements of the target using Gammasphere was carried out using the RadWare \cite{radware} software. The absolute efficiency of Gammasphere operating in singles mode was not measured directly. Instead, an absolute efficiency curve was determined by comparing common radionuclides from both Gammasphere and the single Ge detector analysis. The efficiency curve was constructed using the principal gamma peaks observed in the two analyses and the ratio of the EOB activities (uncorrected for Gammasphere and efficiency corrected for the Ge detector).   The absolute efficiency of Gammasphere calculated in this manner is similar to \cite{torb}.  
 The activities (cumulative yields)  of those nuclides determined by singles counting in Gammasphere after the irradiation are also tabulated in Tables 1 and 2 along with the relevant independent yields.  These cross sections are identified as being from radioactive decay (RD)  measurements in Tables 1 and 2.

The analysis of the in-beam Gammasphere data was also carried out using the RadWare \cite{radware} software. Two $\gamma$-$\gamma$-$\gamma$  histograms, called cubes, were constructed. One cube was constructed using prompt gamma decays recorded during the beam burst (IB), and the other was constructed using delayed gamma decays recorded between the beam bursts (OB). Using the program LEVIT8R, the identification of the reaction products was determined by gating on two low lying gamma transitions in each cube for a given nucleus. The observation of the next higher up transition confirmed the identity of the reaction product and was integrated. This procedure was repeated for all observed three-fold transitions for a given nucleus. The intensities for each three-fold transition were corrected for internal conversion, absolute efficiency, and triple-coincidence efficiency (which was determined using the method outlined in \cite{Cocks1,Cocks2}). All of the individual transition intensities from both the IB cube and OB cube were summed to give the total gamma yields of each reaction product. Once again, absolute cross sections were determined using equations for growth and decay \cite{FKMM}, taking into account the beam intensities.
	
As in the radioactive decay analysis, the yields of ground state and isomeric states were summed to give a total nuclidic cross section. These measured absolute nuclidic cross sections are the ``cumulative" yields for the in-beam data and are represented in Tables 1 and 2 in the Appendix. The cross sections are identified as IB and/or OB in Tables 1 and 2 depending on if the cross section was determined by gamma transitions found in the prompt cube of in-beam burst events (IB), in the delayed cube of out-of-beam burst events (OB), or both. The same procedure to determine independent yields in the decay analysis was then applied to these data to determine independent yields. These independent cross sections are represented in Tables 1 and 2.

\section{Results and Discussion}
\subsection{This Work}
The measured cumulative and independent yields of the projectile-like fragments (PLFs) and target-like fragments (TLFs) from the interaction of E$_{c.m.}$ = 450 MeV $^{136}$Xe with $^{208}$Pb form a large data set (235 yields) to characterize the product distributions from this reaction.  The magnitudes of the measured cross sections range from $\sim$ 2$\mu$b to $\sim$ 75 mb.  The observed PLFs span the region from Z=48 to Z=68 (Xe is Z=54) while the observed TLFs range from Z=70 to Z=88 (Pb is Z = 82).  These yield patterns are presented in Figures 2 and 3.  The observed nuclides are ``north-east" of the projectile and ``south-west" of the target although there are several notable exceptions.  Unknown nuclei cannot be observed using our experimental methods. TLFs with N=126 ranging from $^{206}$Hg to $^{214}$Ra were observed with cross sections ranging from 9 $\mu$b to 3 $\mu$b ($\Delta$Z = -2 to + 6). The N=126 nuclide $^{205}$Au was observed in $\gamma$$\gamma$ coincidence data, but no $\gamma$$\gamma$$\gamma$ coincidences were observed, making it difficult to make a quantitative determination of the yield.  PLFs with N=82 ranging from $^{134}$Te to $^{143}$Pm were observed with cross sections ranging from 261 $\mu$b  to 16 $\mu$b  ($\Delta$Z = -2 to + 7).  

\subsection{Comparison with previous measurements}

Previous measurements of multi-nucleon transfer reactions with $^{208}$Pb include the aforementioned thick target studies of Krolas et al. \cite{krol1,krol2},  the study of Wilson et al. \cite{wilson} of the $^{64}$Ni + $^{208}$Pb reaction where relative $\gamma$-ray intensities of decaying products were reported, the more recent  studies with the velocity filter SHIP of the $^{58,64}$Ni + $^{207}$Pb reaction \cite {heinz, beli} and the study using CORSET of the product yields in the $^{136}$Xe + $^{208}$Pb reaction \cite{kozulin}.  Related studies of the PLFs from the $^{136}$Xe + $^{198}$Pt reaction were reported by the KISS spectrometer group \cite {kim,wata}

The pattern that emerges from these studies is:

(a) The more neutron-rich the projectile, the more neutron-rich the TLFs are \cite{heinz}.

(b) N=126 TLFs ranging from Tl to Ra ($\Delta$Z =-1 to $\Delta$Z = +6) are made in the $^{64}$Ni + $^{207,208}$Pb reactions \cite{heinz,krol1}.

(c) The shapes of the measured isotopic distributions for the $^{64}$Ni + $^{207,208}$Pb reactions using SHIP \cite{beli} and  thick target $\gamma$-ray spectroscopy differ substantially with the distributions from latter measurements being more ``Gaussian-like" in better agreement with theoretical predictions.

(d) In the study of the E$_{lab}$ = 850 MeV $^{136}$Xe + $^{208}$Pb  reaction \cite{kozulin} , the trans-target products  $^{210}$Po, $^{222}$Rn and $^{224}$Ra were observed with production cross sections of 200$\pm$100 $\mu$b, 17$\pm$14 $\mu$b and 2.5 $\pm$2 $\mu$b, respectively.

In Figure 4, we compare the secondary product mass distribution (i.e., not corrected for neutron emission) deduced in this work with those measured previously by \cite{kozulin}.  One should note that the mass distribution measured by \cite{kozulin} excluded quasi-elastic events while that is not possible in this work and is that of the primary fragments before neutron emission.  Also,  the CORSET detector used in \cite{kozulin} has a mass resolution of 7 u while our measurement has a mass resolution of 1 u. Given these qualifications, the agreement between the measurements seems acceptable--which serves as a rough measure of the accuracy of the absolute cross sections measured in each work.  (The measured total reaction cross section in this work is 359 $\pm$ 90 mb while the semi-empirical model of Bass \cite{Bass} would suggest that $\sigma$$_{R}$ is 568 mb and the formalism of \cite{wilcke} predicts $\sigma$$_{R}$ is 496 mb for the reaction of E$_{c.m.}$=452 MeV $^{136}$Xe + $^{208}$Pb .)

In Figure 5,  we compare the Pt (Z=78, $\Delta$Z = -4) isotopic distributions from this work and from the interaction of E$_{lab}$=350 MeV $^{64}$Ni with $^{208}$Pb\cite{krol1}.  As stated above, the more n-rich projectile, $^{136}$Xe, N/Z= 1.52, gives a TLF distribution that peaks at a larger neutron number compared to the less n-rich projectile, $^{64}$Ni, N/Z=1.29.  The distribution from the $^{136}$Xe induced reaction appears to extend out to larger neutron numbers than the $^{64}$Ni induced reaction. 

\subsection{Comparison with phenomenological models}

We shall compare our measurements with the predictions of the GRAZING model and the predictions of the multi-nucleon transfer model of Zagrebaev and Greiner.  To compare our measured cross sections with estimates of these models (which may differ by orders of magnitude), we define a comparison metric \cite{bertsch}, the theory evaluation factor, {\bf tef}.

For each data point, we define 
\begin{equation}
tef_{i}=log\left ( \frac{\sigma _{theory}}{\sigma _{expt}} \right )
\end {equation}
where $\sigma _{theory}$ and $\sigma _{expt}$ are the calculated and measured values of the transfer cross sections.  Then,  the average theory evaluation factor is given by
\begin{equation}
\overline{tef}=\frac{1}{N_{d}}\sum_{i=1}^{N_{d}}tef_{i}
\end{equation}
where N$_{d}$ is the number of data points.  The variance of the average theory evaluation factor is given by 
\begin{equation}
\sigma =\frac{1}{N_{d}}\left ( \sum_{i}\left ( tef_{i}-\overline{tef} \right )^{2} \right)^{1/2}
\end{equation}
Note that {\bf tef} is a logarithmic quantity and theories that have {\bf tef} values differing by  1 or 2 actually differ by orders of magnitude in their reliability.

In Figure 6, we compare the measured Pt ($\Delta$Z=-4) and Rn ($\Delta$Z=+4) isotopic distributions with the calculations of the GRAZING code and those of \cite{z3, kozulin}. There are no published calculations for trans-target nuclei using the Zagrebaev and Greiner formalism for E$_{c.m.}$ = 450 MeV $^{136}$Xe + $^{208}$Pb .  Such calculations exist for E$_{c.m.}$ = 514 MeV \cite{kozulin}.  We have chosen to compare these higher energy calculations with our measurements.  To indicate the effect of the different beam energies, we show GRAZING calculations for E$_{c.m.}$ = 450 MeV (solid line) and E$_{c.m.}$ = 514 MeV (dotted line).  Quantitatively,  the {\bf tef} for GRAZING for the Pt data is -5.0 $\pm$ 0.9 while the {\bf tef} for \cite {z3} is -0.36 $\pm$ 0.15. For the Rn data, the {\bf tef} for GRAZING is -3.3 $\pm$ 0.6 while the {\bf tef} for \cite{kozulin} is -0.69 $\pm$ 0.14.       The GRAZING calculations grossly underestimate the observed cross sections by orders of magnitude and in the case of the Pt isotopic distribution, the mean observed neutron number is overestimated by about 7 neutrons.  {\bf One concludes that for these large proton transfers ($\Delta$Z = $\pm$4) GRAZING  is not a suitable model. } The calculations of Zagrebaev and Greiner do a  better job of estimating the magnitude of the overall transfer cross sections but do not account for  the location of the peaks of the distributions for these large proton transfers.    

In Figure 7, we show the observed values of the cross sections for producing N=126 nuclei in the $^{136}$Xe + $^{208}$Pb reaction.  (We also show the observed cross sections for producing N=126 nuclei in the $^{64}$Ni + $^{208}$Pb reaction.) The point at Z=82 is the target nucleus, $^{208}$Pb, whose yield may include other processes besides multi-nucleon transfer.  Disregarding that point, the other N=126 nuclei form a smooth distribution that peaks at Z=84 (N/Z =1.5).   Only $\Delta$Z =-1 and -2 nuclei were observed on the n-rich side of stability.  The measured cross sections for these nuclei are about two orders of magnitude lower than predicted by the models.  On the n-deficient side of stability the yields disagree with the GRAZING predictions although the measured cross sections for the most neutron deficient nuclei are larger than those predicted by GRAZING.   All in all, this disagreement between models and measurements is not very encouraging for the effort to synthesize neutron-rich N=126 nuclei far from stability.

In Figure 8, we show a contour plot of the measured yields in the $^{136}$Xe + $^{208}$Pb reaction that can be compared to the predictions shown in Figure 1.  The measured cross section contours generally resemble the predicted ones.

In figures 9,10, and 11,  we show the detailed isotopic distributions for TLFs with Z ranging from 74 to 86 (W-Rn) along with the predictions of the GRAZING model \cite{nrvp} and the model of Zagrebaev and Greiner \cite{z1}. (There are no predictions for odd Z nuclei in the Zagrebaev and Greiner formalism.) The trans-target TLF yields are shown in Figure 9.  As in Figure 6, we show the Zagrebaev and Greiner calculations for E$_{c.m.}$ = 514 MeV and GRAZING calculations for 450 MeV (solid line) and 514 MeV (dashed line).The {\bf tef} values for the GRAZING code are as follows: Bi -0.45$\pm$ 0.41, Po -1.14 $\pm$ 0.24, At -1.22 $\pm$ 0.15, and Rn -3.30 $\pm$ 0.64. For Rn, the TEF value for the Zagrebaev and Greiner model \cite{kozulin} is -0.69 $\pm$ 0.14. GRAZING is an adequate model for the small proton transfers(Bi, Po) but fails to describe the larger proton transfer reactions. 

 The ``below-target" TLF yields are shown in Figures 10 and 11.  The {\bf tef} values for the GRAZING code are 0.36 $\pm$ 0.48 (Pb), -0.86 $\pm$ 0.62 (Tl), -1.22 $\pm$ 0.64 (Hg), -5.5$\pm$0.7 (Au), and  -5.0 $\pm$ 0.9 (Pt).  The corresponding values of the {\bf tef} for the calculations of Zagrebaev and Greiner are -0.31 $\pm$ 0.55 (Pb), -0.16 $\pm$0.40 (Hg), +0.54 $\pm$ 0.15 (Pt), -0.23$\pm$ 0.12 (Os), and 0.19 $\pm$ 0.24 (W).
  The GRAZING model is  useful, at best,  for estimating the yields of small transfers ($\Delta$Z$_{TLF}$ = -1 to +2).  In almost all cases,  the model of Zagrebaev and Greiner  gave overall accurate predictions of the transfer product yields, tef $\leq$ 0.5.  However, the predicted peak of the isotopic distributions is  off by 4-6  neutrons in these calculations for large proton transfers, which is troublesome.

\subsection{Correlated nuclidic yields}

Within the limits of the production cross sections, beam intensities and the efficiency of Gammasphere, it is possible to look at the correlations between PLFs and TLFs.  As a demonstration of this, we show in Figure 12 the observed correlated TLFs when the PLF is $^{128}$Te (N/Z = 1.46).  The most probable TLF is $^{205}$Po, which corresponds to  (N/Z=1.44).  This is consistent with the work of Krolas et al. \cite{krol1,krol2} who used the concept of N/Z equilibration to understand the transfer of neutrons and protons amongst the colliding nuclei.  The problem here is that there are 11 unaccounted-for neutrons with a nominal Q value for the charge balanced reaction of -21 MeV.  It would seem to be a useful challenge to theory to predict the observed correlations.

\section{Conclusions}

  To the extent that the system being studied,$^{136}$Xe + $^{208}$Pb, which involves a shell-stabilized projectile and target nucleus is representative of the multi-nucleon transfer process of heavy nuclei, we conclude that:
(a) The GRAZING model is only useful in estimating transfers of $\Delta$Z = -1 to +2.  For larger transfers, the model underestimates the observed cross sections by orders of magnitude.  (b)  The multi-nucleon transfer model of Zagrebaev and Greiner \cite{z1,z2,z3} (and presumably the underlying physics) is remarkably good in predicting the magnitudes of the TLF transfer cross sections for a wide range of transfers ($\Delta$Z = -8 to +4)  (c) The predicted maxima in the TLF transfer product distributions using the model of Zagrebaev and Greiner agree with the observed maxima for $\Delta$Z = -4, -2, and 0.  (For $\Delta$Z = -6 and -8, the observed distributions do not show clear maxima.)  For $\Delta$Z = +4, the observed position of the maximum is 4 neutrons more n-rich than the predicted maximum.  This is consistent with the observation of \cite{kozulin} for the yield of $^{222}$Rn.  (d) The predicted yields of the n-rich N=126 nuclei formed in this reaction exceed the measured yields by orders of magnitude, representing  a significant concern for attempts to synthesize these nuclei. (e) Understanding the correlated  yields of $^{128}$Te and its partners poses a challenge.

From this work alone, we cannot determine  whether the shell-stabilized projectile/target combination studied herein is representative of the larger class of multi-nucleon transfer reactions.  However, based upon the results of this study, a full test of the Zagrebaev and Greiner formalism using the heaviest nuclei would seem to be justified.

\section{Acknowledgements}
This material is based upon work supported in part  by the U.S. Department of Energy, Office of Science, Office of Nuclear Physics under award numbers DE-FG06-97ER41026 (OSU) and DE-FG02-94ER40834 (UMD) and contract numbers DE-AC02-06CH11357 (ANL) and DE-AC02-98CH10886 (BNL).  This research used resources of ANL's ATLAS facility, which is a DOE Office of Science User facility.

\section{Appendix }

\begin{longtable}{|c|c|c|c|}
\caption{Projectile-like fragment cumulative and independent yields for $^{136}$Xe + $^{208}$Pb at E$_{cm}$ = 450 MeV.} \\
\hline
\textbf{Isotope}& \textbf{$\sigma$$_{CY}$ (mb)}& \textbf{$\sigma$$_{IY}$ (mb)}& \textbf{OB/IB/RD} \\
\hline 
\endfirsthead 
\multicolumn{4}{c}{\tablename\ \thetable\ -- \textit{Continued from previous page}} \\
\hline
\textbf{Isotope}& \textbf{$\sigma$$_{CY}$ (mb)}& \textbf{$\sigma$$_{IY}$ (mb)}& \textbf{OB/IB/RD} \\
\hline
\endhead
\hline \multicolumn{4}{c}{\textit{Continued on next page}} \\
\endfoot
\hline
\endlastfoot
$^{116}$Cd& 0.108 $\pm$  0.022 &0.064 $\pm$  0.013  &  OB,IB      \\
$^{118}$Cd& 0.156 $\pm$  0.031 & 0.133 $\pm$  0.027  &    IB       \\
\hline
$^{119}$In& 0.0080$\pm$ 0.0016 &0.0055 $\pm$  0.0011 &    OB       \\
$^{121}$In& 0.0181$\pm$ 0.0036 &0.0168 $\pm$  0.0034 &    OB       \\
\hline
$^{118}$Sn& 0.0134$\pm$ 0.0027 &0.0132 $\pm$  0.0026 &    OB       \\
$^{120}$Sn& 0.237 $\pm$  0.047 &0.098 $\pm$  0.020  &  OB,IB      \\
$^{122}$Sn& 0.443 $\pm$  0.089 & 0.347 $\pm$  0.069  &  OB,IB      \\
$^{123}$Sn& 0.0170$\pm$ 0.0034 &0.0153 $\pm$  0.0031 &  OB,IB      \\
$^{124}$Sn& 0.383 $\pm$  0.077 & 0.367 $\pm$  0.073  &    OB       \\
$^{125}$Sn& 0.0083$\pm$ 0.0017 &0.0083 $\pm$  0.0017 &    OB       \\
$^{126}$Sn& 0.318 $\pm$  0.064 & 0.316 $\pm$  0.063  &  OB,IB      \\
\hline
$^{119}$Sb& 0.0016$\pm$ 0.0003 &0.0016 $\pm$  0.0003 &    OB       \\
$^{121}$Sb& 0.0071$\pm$ 0.0014 &0.0017 $\pm$  0.0003 &    OB       \\
$^{123}$Sb& 0.111 $\pm$  0.022 &0.0583 $\pm$  0.012  &    OB       \\
$^{125}$Sb& 0.189 $\pm$  0.038 & 0.165 $\pm$  0.033  &  OB,IB      \\
$^{126}$Sb& 0.655 $\pm$  0.131  & 0.655 $\pm$   0.131  &    RD       \\
$^{127}$Sb& 0.548 $\pm$  0.11  & 0.51 $\pm$   0.10  & OB,IB,RD    \\
$^{128}$Sb& 0.418 $\pm$  0.084 & 0.396 $\pm$  0.079  &    RD       \\
$^{130}$Sb& 0.0026$\pm$ 0.0005 &0.0026 $\pm$  0.0005 &    OB       \\
\hline
$^{124}$Te& 0.195 $\pm$  0.039 & 0.192 $\pm$  0.038  &  OB,IB      \\
$^{126}$Te&  1.22 $\pm$  0.24  & 0.74 $\pm$   0.15  &  OB,IB      \\
$^{128}$Te&  1.15 $\pm$  0.23  & 1.05  $\pm$   0.21  &  OB,IB      \\
$^{130}$Te&  1.88 $\pm$  0.38  & 1.83  $\pm$   0.37  &  OB,IB      \\
$^{131}$Te&  2.28 $\pm$  0.46  & 1.79  $\pm$   0.36  &  OB,RD      \\
$^{132}$Te&  2.43 $\pm$  0.49  & 2.28  $\pm$   0.46  & OB,IB,RD    \\
$^{134}$Te& 0.261 $\pm$  0.052 & 0.261 $\pm$  0.052  &    OB       \\
\hline
$^{127}$I & 0.0463$\pm$ 0.0093 &0.0223 $\pm$  0.0045 &    IB       \\
$^{128}$I & 0.0849$\pm$  0.017 &0.085 $\pm$  0.017  &    IB       \\
$^{129}$I & 0.934 $\pm$  0.187  & 0.67 $\pm$   0.13  &    IB       \\
$^{130}$I &  3.10 $\pm$  0.62  & 3.10  $\pm$   0.62  &    RD       \\
$^{131}$I &  4.58 $\pm$  0.92  & 3.61  $\pm$   0.72  & OB,IB,RD    \\
$^{132}$I &  5.40 $\pm$   1.08  & 5.3  $\pm$   1.1   &    RD       \\
$^{133}$I &  4.34 $\pm$  0.87  & 3.46  $\pm$   0.69  & OB,IB,RD    \\
$^{135}$I &  2.05 $\pm$  0.41  & 2.04  $\pm$   0.41  &  IB,RD      \\
$^{136}$I & 0.0201$\pm$ 0.0040 &0.0200 $\pm$  0.004 &    IB       \\
\hline
$^{128}$Xe& 0.190 $\pm$  0.038 & 0.188 $\pm$  0.038  &  OB,IB      \\
$^{130}$Xe&  5.72 $\pm$   1.14  & 5.1  $\pm$   1.0   &  OB,IB      \\
$^{132}$Xe&  8.46 $\pm$   1.69  & 5.9  $\pm$   1.2   &  OB,IB      \\
$^{133}$Xe&  18.6 $\pm$   3.7  & 17.2  $\pm$   3.4   & OB,IB,RD    \\
$^{134}$Xe&  10.2 $\pm$   2.0  & 8.6  $\pm$   1.7   &  OB,IB      \\
$^{135}$Xe&  74.0 $\pm$   14.9   & 64  $\pm$    13   & OB,IB,RD    \\
$^{136}$Xe&  31.4 $\pm$   6.3  & 30.4  $\pm$   6.1   &  OB,IB      \\
$^{137}$Xe&  1.48 $\pm$  0.30  & 1.46  $\pm$   0.29  &  OB,IB      \\
$^{138}$Xe&  3.26 $\pm$  0.65  & 3.26  $\pm$   0.65  &    IB       \\
\hline
$^{131}$Cs& 0.212 $\pm$  0.042 & 0.212 $\pm$  0.042  &    IB       \\
$^{132}$Cs&  1.69 $\pm$  0.34  & 1.69  $\pm$   0.34  &  IB,RD      \\
$^{133}$Cs&  1.70 $\pm$  0.34  & 0.75 $\pm$   0.15  &    IB       \\
$^{134}$Cs& 0.829 $\pm$  0.166  & 0.76 $\pm$   0.15  &    IB       \\
$^{136}$Cs&  15.5 $\pm$   3.1  & 15.5  $\pm$   3.1   &    RD       \\
$^{137}$Cs&  4.85 $\pm$  0.97  & 4.43  $\pm$   0.89  &  OB,IB      \\
$^{139}$Cs& 0.473 $\pm$  0.095 & 0.449 $\pm$  0.090  &    IB       \\
$^{141}$Cs& 0.0877$\pm$  0.018 &0.087 $\pm$  0.017  &    IB       \\
\hline
$^{130}$Ba& 0.0057$\pm$ 0.0011 &0.0057 $\pm$  0.0011 &  OB,IB      \\
$^{132}$Ba& 0.0816$\pm$  0.016 &0.0807 $\pm$  0.016  &    IB       \\
$^{134}$Ba& 0.664 $\pm$  0.132  & 0.65 $\pm$   0.13  &  OB,IB      \\
$^{136}$Ba&  1.82 $\pm$  0.36  & 1.59  $\pm$   0.32  &  OB,IB      \\
$^{138}$Ba&  7.97 $\pm$   1.59  & 6.0  $\pm$   1.2   &  OB,IB      \\
$^{139}$Ba&  5.75 $\pm$   1.15  & 4.8  $\pm$   1.0   &    IB       \\
$^{140}$Ba&  3.96 $\pm$  0.79  & 3.30  $\pm$   0.66  & OB,IB,RD    \\
$^{141}$Ba& 0.160 $\pm$  0.032 & 0.160 $\pm$  0.032  &    IB       \\
$^{142}$Ba& 0.163 $\pm$  0.033 & 0.159 $\pm$  0.032  &    IB       \\
$^{143}$Ba& 0.0296$\pm$ 0.0059 &0.0293 $\pm$  0.0059 &    IB       \\
\hline
$^{132}$La& 0.0419$\pm$ 0.0084 &0.0418 $\pm$  0.0084 &    IB       \\
$^{135}$La& 0.352 $\pm$  0.070 & 0.351 $\pm$  0.070  &  OB,IB      \\
$^{136}$La& 0.420 $\pm$  0.084 & 0.410 $\pm$  0.082  &  OB,IB      \\
$^{137}$La&  1.71 $\pm$  0.34  & 1.67  $\pm$   0.33  &  OB,IB      \\
$^{139}$La& 0.952 $\pm$  0.190  & 0.58 $\pm$   0.12  &  OB,IB      \\
$^{140}$La&  2.20 $\pm$  0.44  & 2.18  $\pm$   0.44  &    RD       \\
$^{143}$La& 0.144 $\pm$  0.029 & 0.144 $\pm$  0.029  &    IB       \\
\hline
$^{136}$Ce& 0.0829$\pm$  0.017 &0.083 $\pm$  0.017  &  OB,IB      \\
$^{138}$Ce& 0.351 $\pm$  0.070 & 0.342 $\pm$  0.068  &  OB,IB      \\
$^{139}$Ce&  7.35 $\pm$   1.47  & 6.6  $\pm$   1.3   &  IB,RD      \\
$^{140}$Ce& 0.316 $\pm$  0.063 & 0.282 $\pm$  0.056  &    IB       \\
$^{141}$Ce&  5.65 $\pm$   1.13  & 3.41  $\pm$   0.68  &  IB,RD      \\
$^{142}$Ce&  2.24 $\pm$  0.45  & 1.69  $\pm$   0.34  &    IB       \\
$^{143}$Ce&  1.43 $\pm$  0.29  & 0.89 $\pm$   0.18  &    RD       \\
$^{144}$Ce& 0.598 $\pm$  0.119  & 0.54 $\pm$   0.11  &    IB       \\
$^{145}$Ce& 0.0345$\pm$ 0.0069 &0.0324 $\pm$  0.0065 &    IB       \\
$^{146}$Ce& 0.0956$\pm$  0.0191 &0.096 $\pm$  0.019  &    IB       \\
\hline
$^{139}$Pr& 0.0284$\pm$ 0.0057 &0.0280 $\pm$  0.0056 &    IB       \\
$^{141}$Pr& 0.342 $\pm$  0.068 & 0.326 $\pm$  0.065  &    IB       \\
$^{142}$Pr&  1.11 $\pm$  0.22  & 1.11  $\pm$   0.22  &    RD       \\
\hline
$^{140}$Nd& 0.0289$\pm$ 0.0058 &0.0287 $\pm$  0.0057 &    OB       \\
$^{142}$Nd&  2.47 $\pm$  0.49  & 2.39  $\pm$   0.48  &  OB,IB      \\
$^{143}$Nd&  13.0 $\pm$   2.6  & 3.41  $\pm$   0.68  &  OB,IB      \\
$^{144}$Nd&  7.45 $\pm$   1.49  & 2.79  $\pm$   0.56  &  OB,IB      \\
$^{145}$Nd& 0.245 $\pm$  0.049 & 0.162 $\pm$  0.032  &    IB       \\
$^{146}$Nd& 0.311 $\pm$  0.062 & 0.250 $\pm$  0.050  &    IB       \\
$^{147}$Nd& 0.228 $\pm$  0.046 & 0.199 $\pm$  0.040  &    IB       \\
$^{148}$Nd& 0.276 $\pm$  0.055 & 0.253 $\pm$  0.051  &    IB       \\
$^{149}$Nd& 0.134 $\pm$  0.027 & 0.131 $\pm$  0.026  &    IB       \\
\hline
$^{142}$Pm& 0.158 $\pm$  0.032 & 0.157 $\pm$  0.032  &    IB       \\
$^{143}$Pm& 0.0156$\pm$ 0.0031 &0.0153 $\pm$  0.0031 &    IB       \\
$^{145}$Pm& 0.253 $\pm$  0.051 & 0.239 $\pm$  0.048  &    IB       \\
$^{147}$Pm& 0.147 $\pm$  0.029 &0.0934 $\pm$  0.019  &    IB       \\
$^{149}$Pm& 0.124 $\pm$  0.025 & 0.114 $\pm$  0.023  &    IB       \\
\hline
$^{145}$Sm& 0.0516$\pm$  0.0103 &0.051 $\pm$  0.010  &    IB       \\
$^{146}$Sm&  1.59 $\pm$  0.32  & 1.52  $\pm$   0.30  &    IB       \\
$^{147}$Sm&  2.05 $\pm$  0.41  & 1.88  $\pm$   0.38  &    IB       \\
$^{148}$Sm&  1.11 $\pm$  0.22  & 0.51 $\pm$   0.10  &    IB       \\
$^{149}$Sm& 0.0557$\pm$  0.0111 &0.0434 $\pm$  0.0087 &    IB       \\
$^{150}$Sm& 0.224 $\pm$  0.045 & 0.194 $\pm$  0.039  &  OB,IB      \\
$^{151}$Sm& 0.0348$\pm$ 0.0070 &0.0322 $\pm$  0.0064 &    IB       \\
$^{152}$Sm& 0.0586$\pm$  0.0111 &0.054 $\pm$  0.011  &  OB,IB      \\
$^{154}$Sm& 0.203 $\pm$  0.041 & 0.191 $\pm$  0.038  &    IB       \\
\hline
$^{147}$Eu& 0.248 $\pm$  0.050 & 0.244 $\pm$  0.049  &    IB       \\
$^{149}$Eu& 0.146 $\pm$  0.029 & 0.145 $\pm$  0.029  &    IB       \\
$^{151}$Eu& 0.0305$\pm$ 0.0061 &0.0305 $\pm$  0.0061 &    IB       \\
\hline
$^{152}$Gd& 0.0436$\pm$ 0.0087 &0.0402 $\pm$  0.0080 &  OB,IB      \\
$^{154}$Gd& 0.0250$\pm$ 0.0050 &0.0225 $\pm$  0.0045 &  OB,IB      \\
$^{156}$Gd& 0.0179$\pm$ 0.0036 &0.0115 $\pm$  0.0023 &    IB       \\
\hline
$^{156}$Dy& 0.0468$\pm$ 0.0094 &0.0453 $\pm$  0.0091 &    IB       \\
$^{158}$Dy& 0.0207$\pm$ 0.0041 &0.0185 $\pm$  0.0037 &    IB       \\
$^{160}$Dy& 0.0173$\pm$ 0.0035 &0.0130 $\pm$  0.0026 &    IB       \\
$^{162}$Dy& 0.0418$\pm$ 0.0084 &0.0344 $\pm$  0.0069 &    IB       \\
$^{164}$Dy& 0.0529$\pm$  0.0106 &0.049 $\pm$  0.010  &    IB       \\
\hline
$^{160}$Er& 0.131 $\pm$  0.026 & 0.124 $\pm$  0.025  &  IB,RD      \\
$^{161}$Er& 0.0139$\pm$ 0.0028 &0.0135 $\pm$  0.0027 &    IB       \\
\hline
\end{longtable}

\begin{longtable}{|c|c|c|c|}
\caption{Target-like fragment cumulative and independent yields for $^{136}$Xe + $^{208}$Pb at E$_{cm}$ = 450 MeV.} \\
\hline
\textbf{Isotope}& \textbf{$\sigma$$_{CY}$ (mb)}& \textbf{$\sigma$$_{IY}$ (mb)}& \textbf{OB/IB/RD} \\
\hline 
\endfirsthead 
\multicolumn{4}{c}{\tablename\ \thetable\ -- \textit{Continued from previous page}} \\
\hline
\textbf{Isotope}& \textbf{$\sigma$$_{CY}$ (mb)}& \textbf{$\sigma$$_{IY}$ (mb)}& \textbf{OB/IB/RD} \\
\hline
\endhead
\hline \multicolumn{4}{c}{\textit{Continued on next page}} \\
\endfoot
\hline
\endlastfoot
$^{176}$Yb& 0.0069$\pm$ 0.0014 &0.0067 $\pm$  0.0013 &    OB       \\
\hline
$^{176}$Hf& 0.0228$\pm$ 0.0046 &0.0185 $\pm$  0.0037 &    OB       \\
$^{178}$Hf& 0.0781$\pm$  0.0156 &0.068 $\pm$  0.014  &    OB       \\
$^{180}$Hf& 0.482 $\pm$  0.096 & 0.474 $\pm$  0.095  &  OB,RD      \\
$^{181}$Hf& 0.0049$\pm$ 0.0010 &0.0045 $\pm$  0.0009 &    OB       \\
$^{182}$Hf& 0.0117$\pm$ 0.0023 &0.0112 $\pm$  0.0022 &    OB       \\
\hline
$^{179}$Ta& 0.0245$\pm$ 0.0049 &0.0156 $\pm$  0.0031 &    OB       \\
$^{181}$Ta& 0.0247$\pm$ 0.0049 &0.0187 $\pm$  0.0037 &    OB       \\
\hline
$^{176}$W & 0.0179$\pm$ 0.0036 &0.0174 $\pm$  0.0035 &    IB       \\
$^{180}$W & 0.0461$\pm$ 0.0092 &0.0417 $\pm$  0.0083 &  OB,IB      \\
$^{182}$W & 0.0289$\pm$ 0.0058 &0.0237 $\pm$  0.0047 &  OB,IB      \\
$^{184}$W & 0.0428$\pm$ 0.0086 &0.0383 $\pm$  0.0077 &  OB,IB      \\
$^{186}$W & 0.0286$\pm$ 0.0057 &0.0261 $\pm$  0.0052 &  OB,IB      \\
$^{187}$W & 0.0965$\pm$  0.019 &0.097 $\pm$  0.019  &    IB       \\
\hline
$^{179}$Re& 0.0035$\pm$ 0.0007 &0.0034 $\pm$  0.0007 &    OB       \\
$^{185}$Re& 0.0246$\pm$ 0.0049 &0.0154 $\pm$  0.0031 &    OB       \\
$^{187}$Re& 0.0112$\pm$ 0.0022 &0.0100 $\pm$  0.0020 &    OB       \\
\hline
$^{186}$Os& 0.0106$\pm$ 0.0021 &0.0099 $\pm$  0.0020 &  OB,IB      \\
$^{188}$Os& 0.131 $\pm$  0.026 & 0.103 $\pm$  0.021  &  OB,IB      \\
$^{190}$Os&  1.20 $\pm$  0.24  & 0.56 $\pm$   0.11  &  OB,IB      \\
$^{191}$Os& 0.124 $\pm$  0.025 &0.080 $\pm$  0.016  &  OB,IB      \\
$^{192}$Os& 0.144 $\pm$  0.029 & 0.134 $\pm$  0.027  &  OB,IB      \\
$^{194}$Os& 0.0789$\pm$  0.016 &0.0789 $\pm$  0.0158  &  OB,IB      \\
$^{197}$Os& 0.0035$\pm$ 0.0007 &0.0035 $\pm$  0.0007 &    OB       \\
\hline
$^{188}$Ir& 0.116 $\pm$  0.023 & 0.115 $\pm$  0.023  &    RD       \\
$^{190}$Ir&  5.66 $\pm$   1.13  & 3.58  $\pm$   0.72  &    RD       \\
$^{192}$Ir& 0.322 $\pm$  0.064 & 0.322 $\pm$  0.064  &    RD       \\
\hline
$^{190}$Pt& 0.124 $\pm$  0.025 & 0.116 $\pm$  0.023  &  OB,IB      \\
$^{191}$Pt& 0.125 $\pm$  0.025 & 0.115 $\pm$  0.023  &    IB       \\
$^{192}$Pt& 0.473 $\pm$  0.095 & 0.427 $\pm$  0.085  &  OB,IB      \\
$^{194}$Pt&  1.60 $\pm$  0.32  & 1.28  $\pm$   0.26  &  OB,IB      \\
$^{196}$Pt&  4.97 $\pm$  0.99  & 4.21  $\pm$   0.84  &  OB,IB      \\
$^{197}$Pt&  4.81 $\pm$  0.96  & 3.51  $\pm$   0.70  & OB,IB,RD    \\
$^{198}$Pt& 0.666 $\pm$  0.13  & 0.63 $\pm$   0.13  &  OB,IB      \\
$^{200}$Pt& 0.728 $\pm$  0.15  & 0.73 $\pm$   0.15  & OB,IB,RD    \\
$^{201}$Pt& 0.0714$\pm$  0.014 &0.0714 $\pm$  0.014  &  OB,IB      \\
$^{202}$Pt& 0.178 $\pm$  0.036 & 0.178 $\pm$  0.036  &    OB       \\
\hline
$^{191}$Au& 0.514 $\pm$  0.103  & 0.469 $\pm$  0.094  &  OB,RD      \\
$^{192}$Au& 0.168 $\pm$  0.034 & 0.155 $\pm$  0.031  &    RD       \\
$^{193}$Au&  1.39 $\pm$  0.28  & 1.27  $\pm$   0.21  &    RD       \\
$^{194}$Au&  1.31 $\pm$  0.26  & 1.26  $\pm$   0.25  &    RD       \\
$^{196}$Au&  4.05 $\pm$  0.81  & 3.34  $\pm$   0.67  &    RD       \\
$^{198}$Au&  2.81 $\pm$  0.56  & 2.77  $\pm$   0.55  &    RD       \\
$^{199}$Au&  8.57 $\pm$   1.7  & 6.23  $\pm$   1.25   &    RD       \\
\hline
$^{194}$Hg& 0.457 $\pm$  0.091 & 0.446 $\pm$  0.089  &  OB,IB      \\
$^{196}$Hg& 0.897 $\pm$  0.179  & 0.847 $\pm$   0.169  &  OB,IB      \\
$^{198}$Hg&  2.81 $\pm$  0.56  & 2.39  $\pm$   0.48  &  OB,IB      \\
$^{200}$Hg&  7.49 $\pm$   1.50  & 6.07  $\pm$   1.21   &  OB,IB      \\
$^{202}$Hg&  2.66 $\pm$  0.53  & 2.10  $\pm$   0.42  &  OB,IB      \\
$^{203}$Hg&  6.61 $\pm$   1.32  & 5.30  $\pm$   1.06   & OB,IB,RD    \\
$^{204}$Hg&  8.68 $\pm$   1.74  & 6.75  $\pm$   1.35   &  OB,IB      \\
$^{205}$Hg&  5.93 $\pm$   1.19  & 4.95  $\pm$   0.99   &    OB       \\
$^{206}$Hg& 0.0093$\pm$ 0.0019 &0.0093 $\pm$  0.0019 &    OB       \\
$^{208}$Hg& 0.0341$\pm$ 0.0068 &0.0341 $\pm$  0.0068 &    OB       \\
\hline
$^{196}$Tl& 0.281 $\pm$  0.056 & 0.283 $\pm$  0.057  &    RD       \\
$^{197}$Tl& 0.564 $\pm$  0.11  & 0.521 $\pm$   0.104  &  IB,RD      \\
$^{198}$Tl& 0.386 $\pm$  0.077 & 0.342 $\pm$  0.068  &    RD       \\
$^{199}$Tl&  1.46 $\pm$  0.29  & 1.30  $\pm$   0.26  &  IB,RD      \\
$^{201}$Tl&  8.95 $\pm$   1.79  & 8.14  $\pm$   1.63   & OB,IB,RD    \\
$^{202}$Tl& 0.754 $\pm$  0.15  & 0.754 $\pm$   0.151  &    IB       \\
$^{203}$Tl& 0.360 $\pm$  0.072 & 0.239 $\pm$  0.048  &    OB       \\
$^{204}$Tl&  1.68 $\pm$  0.34  & 1.68  $\pm$   0.34  &  OB,IB      \\
$^{205}$Tl&  9.93 $\pm$   1.99  & 6.39  $\pm$   1.28   &  OB,IB      \\
$^{206}$Tl&  10.5 $\pm$   2.1  & 9.03  $\pm$   1.81   &  OB,IB      \\
$^{207}$Tl& 0.145 $\pm$  0.029 & 0.129 $\pm$  0.026  &  OB,IB      \\
\hline
$^{198}$Pb& 0.0236$\pm$ 0.0047 &0.0233 $\pm$  0.0047 &    OB       \\
$^{201}$Pb&  1.56 $\pm$  0.31  & 1.41  $\pm$   0.28  & OB,IB,RD    \\
$^{202}$Pb&  4.84 $\pm$  0.97  & 4.55  $\pm$   0.91  & OB,IB,RD    \\
$^{203}$Pb&  6.41 $\pm$   1.28  & 5.95  $\pm$   1.19   &  OB,RD      \\
$^{204}$Pb&  5.66 $\pm$   1.13  & 4.30  $\pm$   0.86  & OB,IB,RD    \\
$^{206}$Pb&  17.2 $\pm$   3.4  & 8.59  $\pm$   1.72   &  OB,IB      \\
$^{207}$Pb& 0.304 $\pm$  0.061 & 0.191 $\pm$  0.038  &    IB       \\
$^{208}$Pb&  25.8 $\pm$   5.2  & 20.6  $\pm$   4.1   &  OB,IB      \\
$^{209}$Pb&  2.20 $\pm$  0.44  & 1.89  $\pm$   0.38  &  OB,IB      \\
$^{210}$Pb& 0.345 $\pm$  0.069 & 0.312 $\pm$  0.062  &  OB,IB      \\
$^{211}$Pb& 0.760 $\pm$  0.152  & 0.760 $\pm$   0.152  &  OB,IB      \\
\hline
$^{199}$Bi& 0.0062$\pm$ 0.0012 &0.0062 $\pm$  0.0012 &    OB       \\
$^{201}$Bi& 0.0299$\pm$ 0.0060 &0.0296 $\pm$  0.0059 &    OB       \\
$^{202}$Bi& 0.0073$\pm$ 0.0015 &0.0071 $\pm$  0.0014 &    OB       \\
$^{203}$Bi& 0.396 $\pm$  0.079 & 0.349 $\pm$  0.070  &  OB,RD      \\
$^{204}$Bi&  2.07 $\pm$  0.41  & 1.93  $\pm$   0.39  &  OB,RD      \\
$^{205}$Bi&  2.17 $\pm$  0.43  & 1.90  $\pm$   0.38  & OB,IB,RD    \\
$^{206}$Bi&  2.97 $\pm$  0.59  & 2.95  $\pm$   0.59  & OB,IB,RD    \\
$^{207}$Bi&  2.60 $\pm$  0.52  & 2.47  $\pm$   0.49  &  OB,IB      \\
$^{209}$Bi& 0.439 $\pm$  0.088 & 0.339 $\pm$  0.068  &  OB,IB      \\
$^{211}$Bi& 0.130 $\pm$  0.026 & 0.127 $\pm$  0.025  &  OB,IB      \\
\hline
$^{202}$Po& 0.0093$\pm$ 0.0019 &0.0092 $\pm$  0.0018 &    OB       \\
$^{204}$Po& 0.104 $\pm$  0.021 & 0.103 $\pm$  0.021  &  OB,IB      \\
$^{205}$Po& 0.124 $\pm$  0.025 & 0.122 $\pm$  0.024  &  OB,IB      \\
$^{206}$Po& 0.815 $\pm$  0.163  & 0.755 $\pm$   0.151  & OB,IB,RD    \\
$^{207}$Po& 0.818 $\pm$  0.164  & 0.817 $\pm$   0.163  & OB,IB,RD    \\
$^{208}$Po&  5.02 $\pm$   1.0  & 4.58  $\pm$   0.92  &  OB,IB      \\
$^{209}$Po& 0.590 $\pm$  0.118  & 0.532 $\pm$   0.106  &  OB,IB      \\
$^{210}$Po&  2.20 $\pm$  0.44  & 1.19  $\pm$   0.24  &  OB,IB      \\
$^{212}$Po& 0.469 $\pm$  0.094 & 0.464 $\pm$  0.093  &  OB,IB      \\
$^{213}$Po& 0.193 $\pm$  0.039 & 0.190 $\pm$  0.038  &  OB,IB      \\
$^{214}$Po& 0.0767$\pm$  0.015 &0.0754 $\pm$  0.0151  &  OB,IB      \\
\hline
$^{207}$At& 0.0102$\pm$ 0.0020 &0.0101 $\pm$  0.0020 &    OB       \\
$^{208}$At& 0.0367$\pm$ 0.0073 &0.0353 $\pm$  0.0071 &    OB       \\
$^{209}$At& 0.635 $\pm$  0.131  & 0.569 $\pm$   0.114  &  OB,RD      \\
$^{210}$At& 0.989 $\pm$  0.198  & 0.894 $\pm$   0.179  &  OB,RD      \\
$^{211}$At& 0.467 $\pm$  0.093 & 0.451 $\pm$  0.090  &  OB,IB      \\
$^{213}$At& 0.384 $\pm$  0.077 & 0.384 $\pm$  0.077  &  OB,IB      \\
\hline
$^{210}$Rn& 0.0700$\pm$  0.014 &0.0678 $\pm$  0.0136  &    OB       \\
$^{211}$Rn& 0.368 $\pm$  0.074 & 0.350 $\pm$  0.070  &  OB,RD      \\
$^{212}$Rn& 0.537 $\pm$  0.107  & 0.515 $\pm$   0.103  &  OB,IB      \\
$^{213}$Rn& 0.166 $\pm$  0.033 & 0.146 $\pm$  0.029  &    OB       \\
$^{214}$Rn& 0.324 $\pm$  0.065 & 0.324 $\pm$  0.065  &  OB,IB      \\
$^{215}$Rn& 0.343 $\pm$  0.069 & 0.343 $\pm$  0.069  &  OB,IB      \\
$^{216}$Rn& 0.112 $\pm$  0.022 & 0.112 $\pm$  0.022  &    IB       \\
$^{218}$Rn& 0.0214$\pm$ 0.0043 &0.0191 $\pm$  0.0038 &    IB       \\
\hline
$^{211}$Fr& 0.0030$\pm$ 0.0006 &0.0030 $\pm$  0.0006 &    OB       \\
$^{212}$Fr& 0.0512$\pm$  0.0102 &0.0512 $\pm$  0.0102  &    OB       \\
$^{213}$Fr& 0.0402$\pm$ 0.0080 &0.0402 $\pm$  0.0080 &    OB       \\
$^{215}$Fr& 0.0167$\pm$ 0.0033 &0.0167 $\pm$  0.0033 &    OB       \\
$^{216}$Fr& 0.0057$\pm$ 0.0011 &0.0057 $\pm$  0.0011 &    OB       \\
\hline
$^{214}$Ra& 0.0034$\pm$ 0.0007 &0.0033 $\pm$  0.0007 &    OB       \\
\hline
\end{longtable}

%\subsection{}
%\subsubsection{}

% If in two-column mode, this environment will change to single-column
% format so that long equations can be displayed. Use
%\begin{widetext}
%\end{widetext}

% figures should be put into the text as floats.
% Use the graphics or graphicx packages (distributed with LaTeX2e)
% and the \includegraphics macro defined in those packages.
% See the LaTeX Graphics Companion by Michel Goosens, Sebastian Rahtz,
% entire page. There is no need to do explicit centering.

% \begin{figure}
% \end{figure}

% Surround figure environment with turnpage environment for landscape
% figure
% \begin{figure}
% \end{figure}

\begin{figure}[h]
\begin{minipage}{35pc}
\begin{center}
\includegraphics[width=35pc]{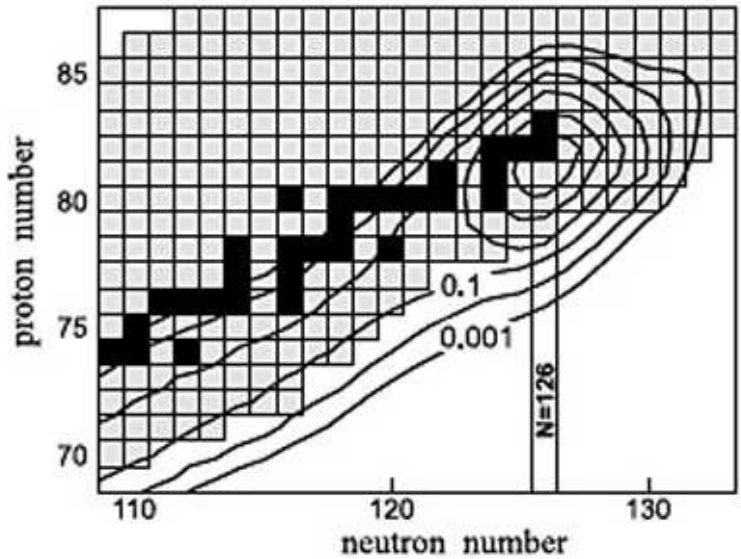}
\end{center}
\caption{\label{fig1}Predicted yields of heavy nuclei in collisions of $^{136}$Xe with $^{208}$Pb at E$_{c.m.}$ = 450 MeV.  From Ref. \cite{z1}}
\end{minipage} 
\end{figure}

\begin{figure}[h]
\begin{center}
\includegraphics[width=40pc]{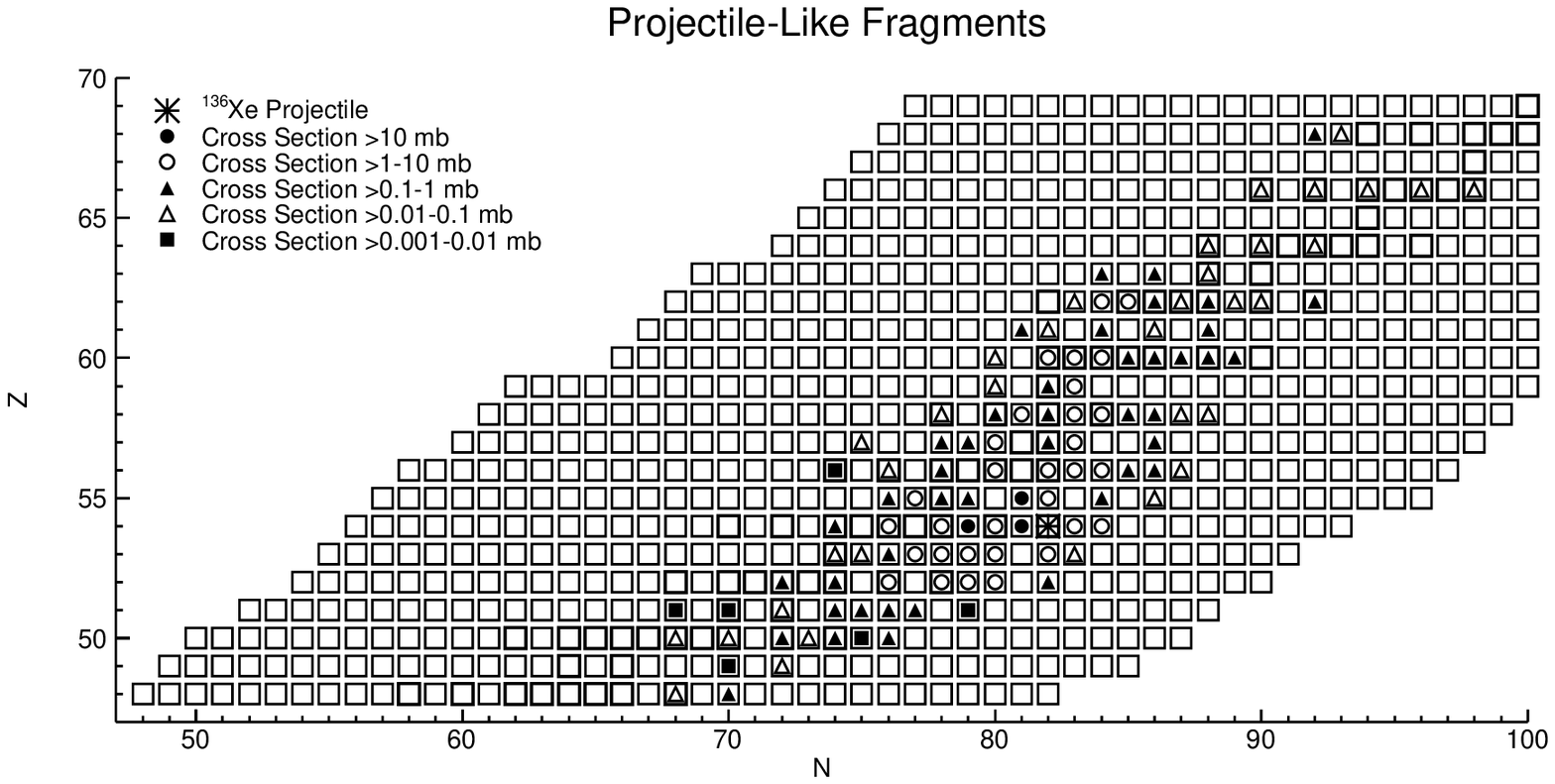}
\end{center}
\caption{\label{fig2}Distribution of PLFs produced in the reaction of E$_{c.m.}$=450 MeV $^{136}$Xe with a thick $^{208}$Pb target.}
\end{figure}

\begin{figure}[h]
\begin{center}
\includegraphics[width=35pc]{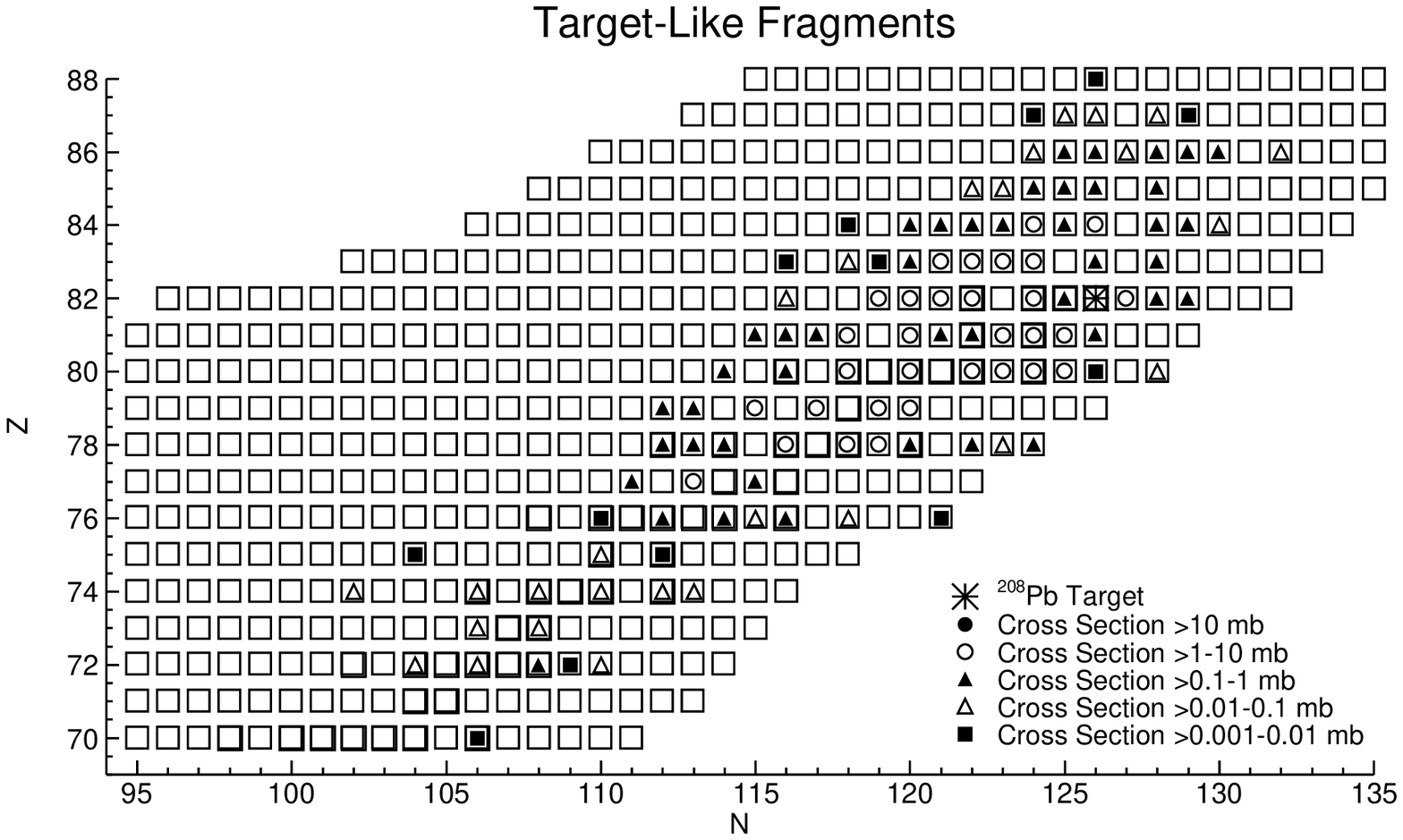}
\end{center}
\caption{\label{fig3}Distribution of TLFs produced in the reaction of E$_{c.m.}$=450 MeV $^{136}$Xe with a thick $^{208}$Pb target.}
\end{figure}

\begin{figure}[h]
\begin{center}
\includegraphics[width=35pc]{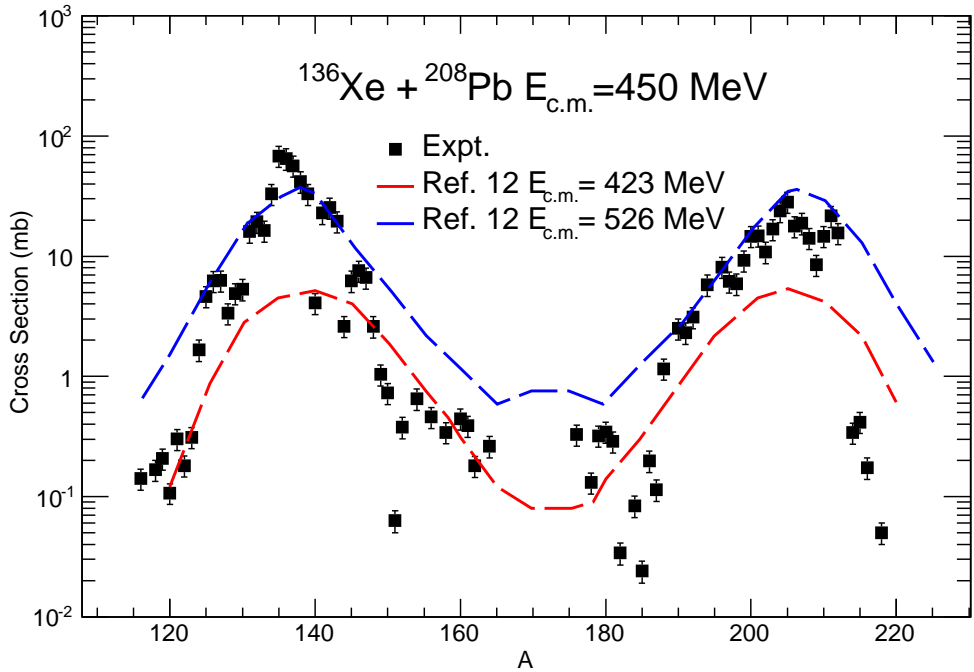}
\end{center}
\caption{\label{fig4} (Color online) Mass distribution of the {\bf secondary} products produced in the reaction of E$_{c.m.}$=450 MeV $^{136}$Xe with a thick $^{208}$Pb target (this work).  Also shown is the reported \cite{kozulin} mass distribution of the {\bf primary} fragments with total kinetic energy loss $\geq$ 40 MeV in the interaction of E$_{c.m.}$ = 423 MeV and E$_{c.m.}$ = 526  MeV $^{136}$Xe with $^{208}$Pb.}
\end{figure}

\begin{figure}[h]
\begin{center}
\includegraphics[width=35pc]{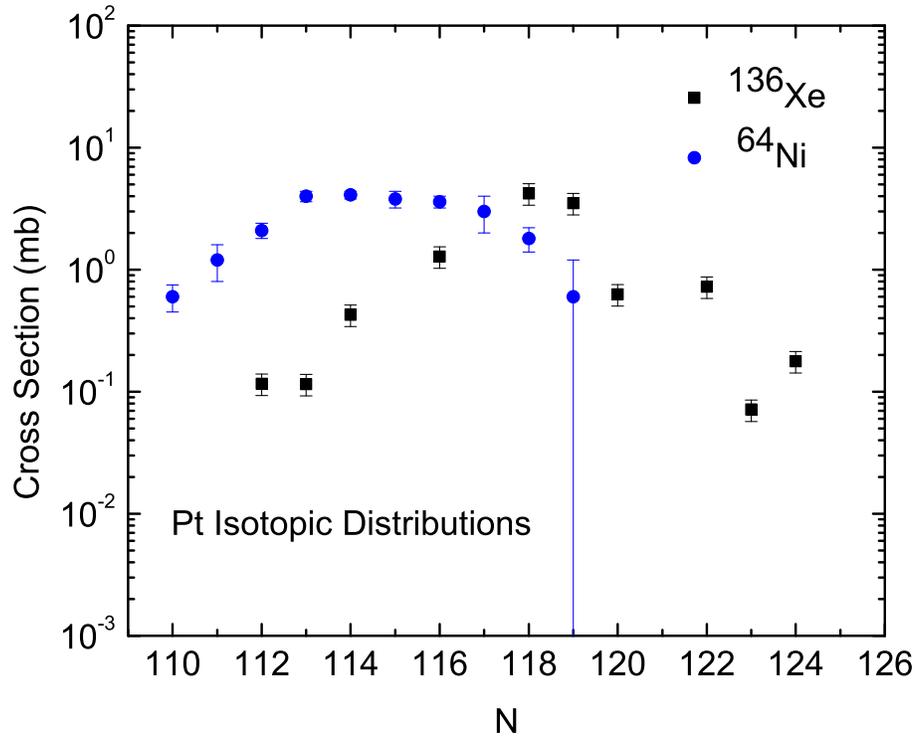}
\end{center}
\caption{\label{fig5}(Color On-line)Distribution of Pt isotopes produced in the reaction of E$_{c.m.}$=450 MeV $^{136}$Xe and E$_{c.m.}$=254 MeV $^{64}$Ni with a thick $^{208}$Pb target.}
\end{figure}

\begin{figure}[h]
\begin{center}
\includegraphics[width=35pc]{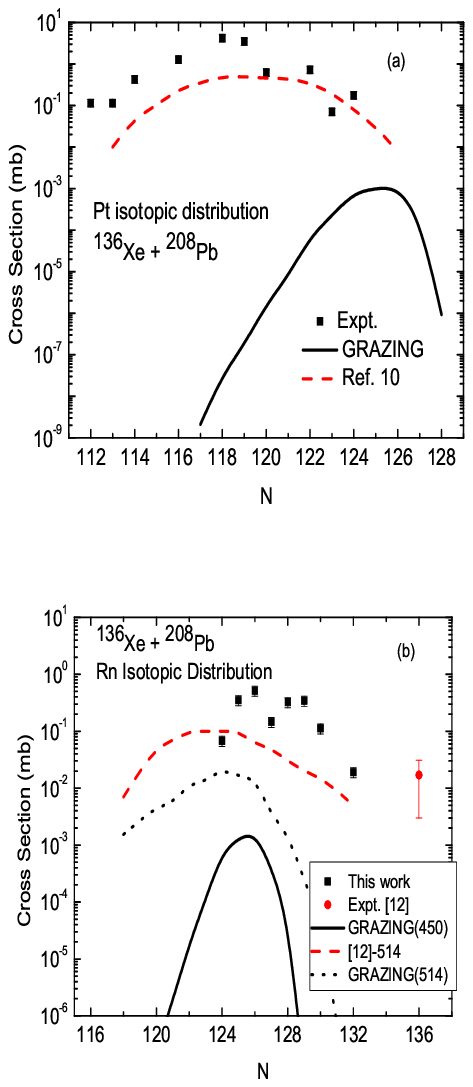}
\end{center}
\caption{\label{fig6}(Color online)(a)Distribution of Pt isotopes produced in the reaction of E$_{c.m.}$=450 MeV $^{136}$Xe with a thick $^{208}$Pb target.(b) Distribution of Rn isotopes produced in the reaction of E$_{c.m.}$=450 MeV $^{136}$Xe  with a thick $^{208}$Pb target.  Also shown are the measurements of  \cite{kozulin} and the predictions of the GRAZING code and \cite{z3,kozulin}. The beam energy used in \cite{kozulin} was E$_{c.m.}$ = 514 MeV.}.  See text for details.
\end{figure}

%\begin{figure}[h]
%\begin{center}
%\includegraphics[width=35pc]{RN.eps}
%\end{center}
%\caption{\label{fig6b}Distribution of Rn isotopes produced in the reaction of E$_{c.m.}$=450 MeV $^{136}$Xe  with a thick $^{208}$Pb target.  Also shown are the measurements and predictions of \cite{kozulin}}
%\end{figure}

\begin{figure}[h]
\begin{center}
\includegraphics[width=35pc]{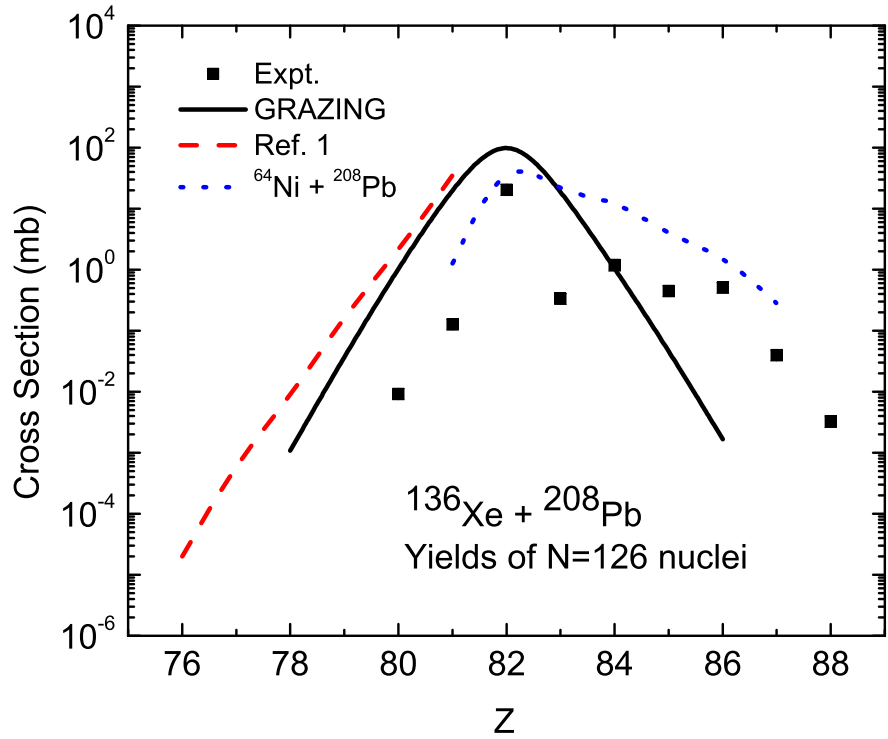}
\end{center}
\caption{\label{fig7}(Color online)Distribution of N=126 nuclides produced in the reaction of E$_{c.m.}$=450 MeV $^{136}$Xe and E$_{c.m.}$=254 MeV $^{64}$Ni with a thick $^{208}$Pb target.}
\end{figure}

\begin{figure}[h]
\begin{minipage}{35pc}
\begin{center}
\includegraphics[width=35pc]{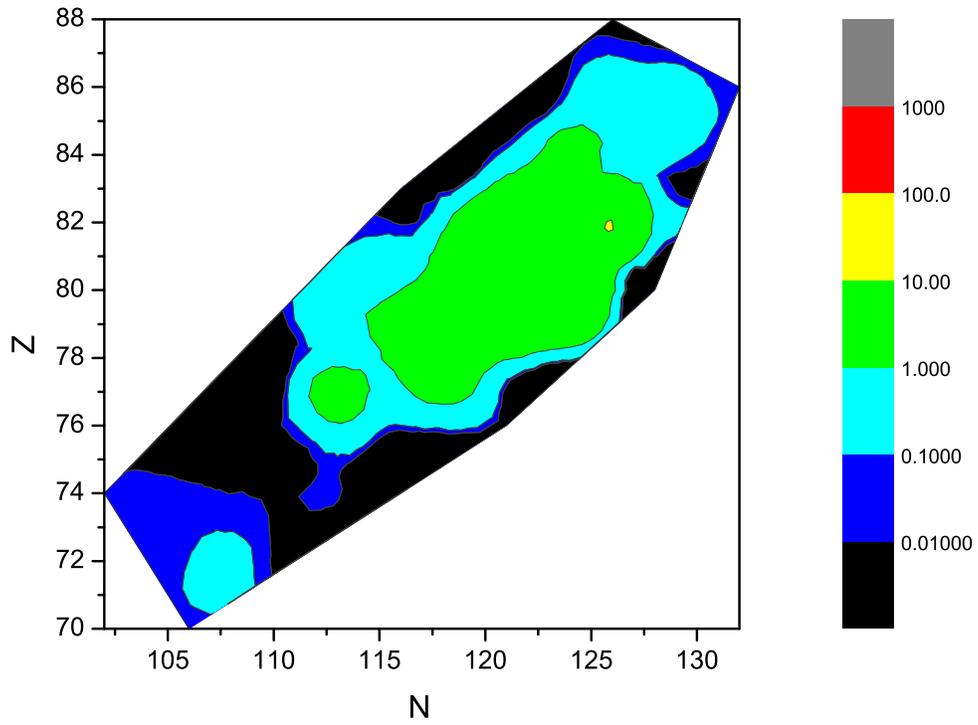}
\end{center}
\caption{\label{fig8}(Color online)Contour plot of measured yields of heavy nuclei in collisions of $^{136}$Xe with $^{208}$Pb at E$_{c.m.}$ = 450 MeV. }
\end{minipage} 
\end{figure}

\begin{figure}[h]
\begin{center}
\includegraphics[width=50pc]{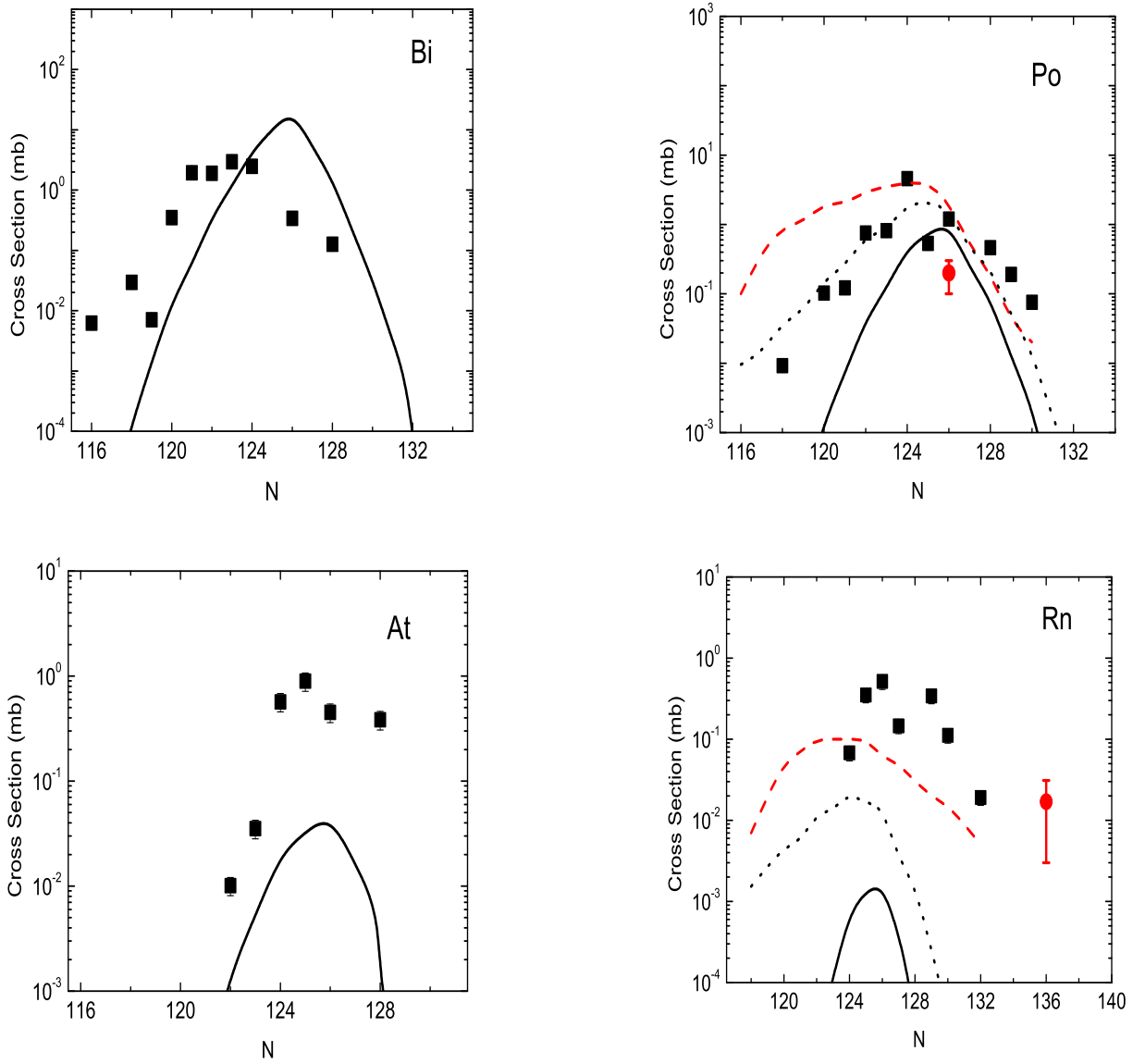}
\end{center}
\caption{\label{fig9}(Color online)Measured distributions of trans-target TLFs compared to the predictions of GRAZING (solid line, dotted line) and \cite{kozulin}(dashed line).A data point from \cite{kozulin} is also shown. The beam energy used in \cite{kozulin} was E$_{c.m.}$ = 514 MeV.}  See text for details.
\end{figure}

\begin{figure}[h]
\begin{center}
\includegraphics[width=50pc]{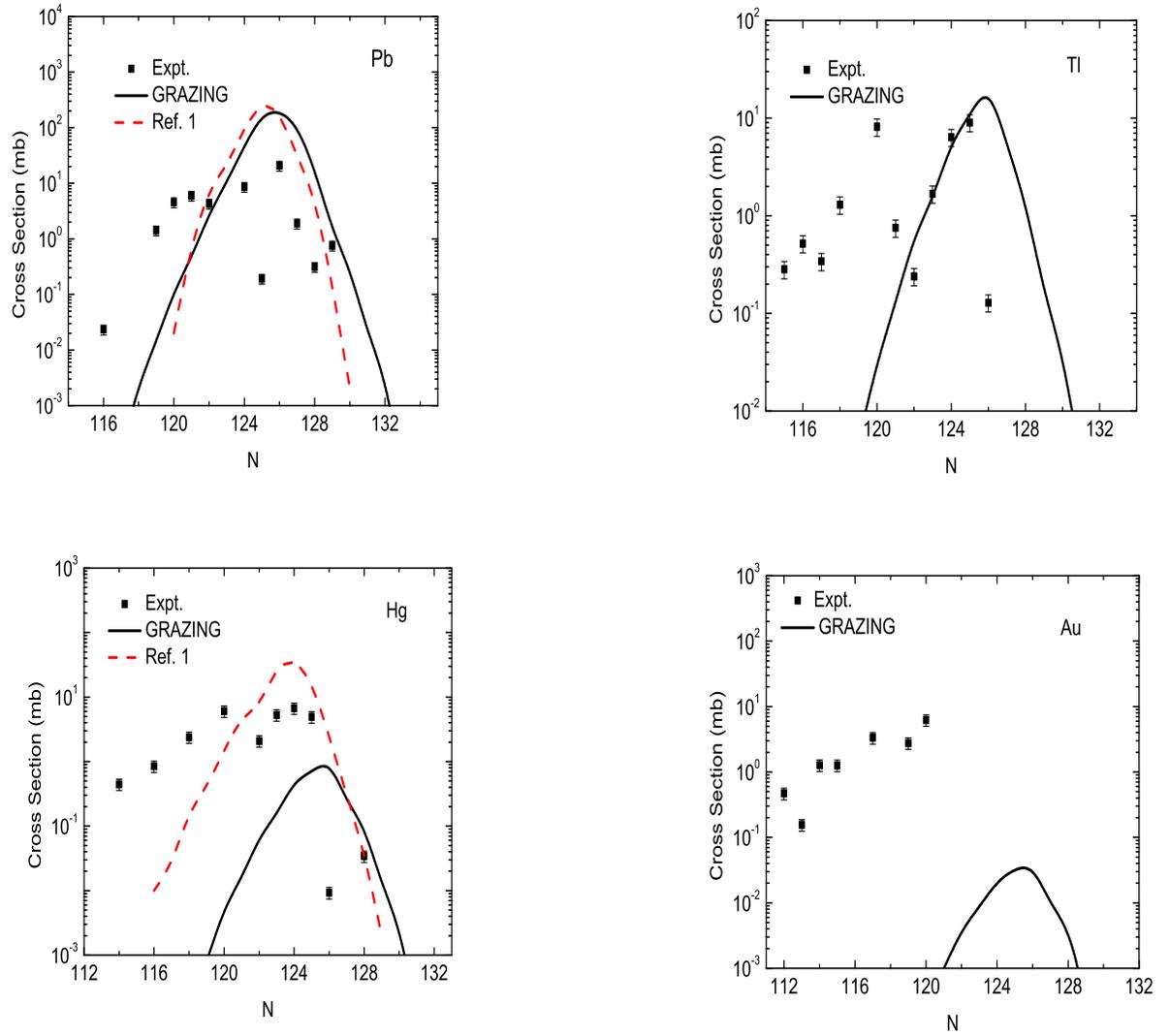}
\end{center}
\caption{\label{fig10}(Color online)Measured distributions of below-target TLFs compared to the predictions of GRAZING (solid line) and Ref. \cite{z1}, dashed line.}
\end{figure}

\begin{figure}[h]
\begin{center}
\includegraphics[width=45pc]{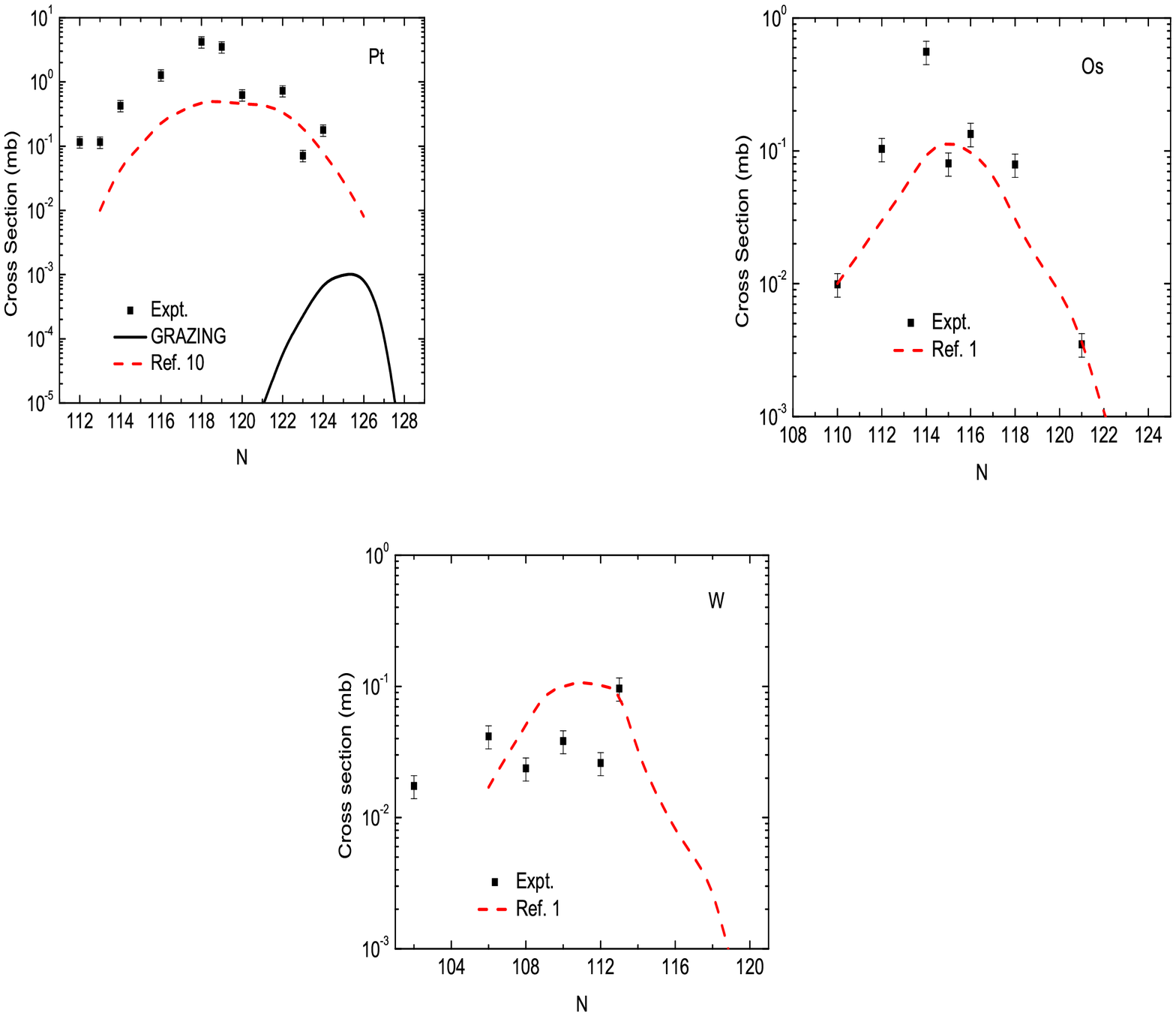}
\end{center}
\caption{\label{fig11}Measured distributions of below-target TLFs compared to the predictions of GRAZING (solid line) and Ref. \cite{z1,z3}, dashed line.}
\end{figure}

\begin{figure}[h]
\begin{center}
\includegraphics[width=35pc]{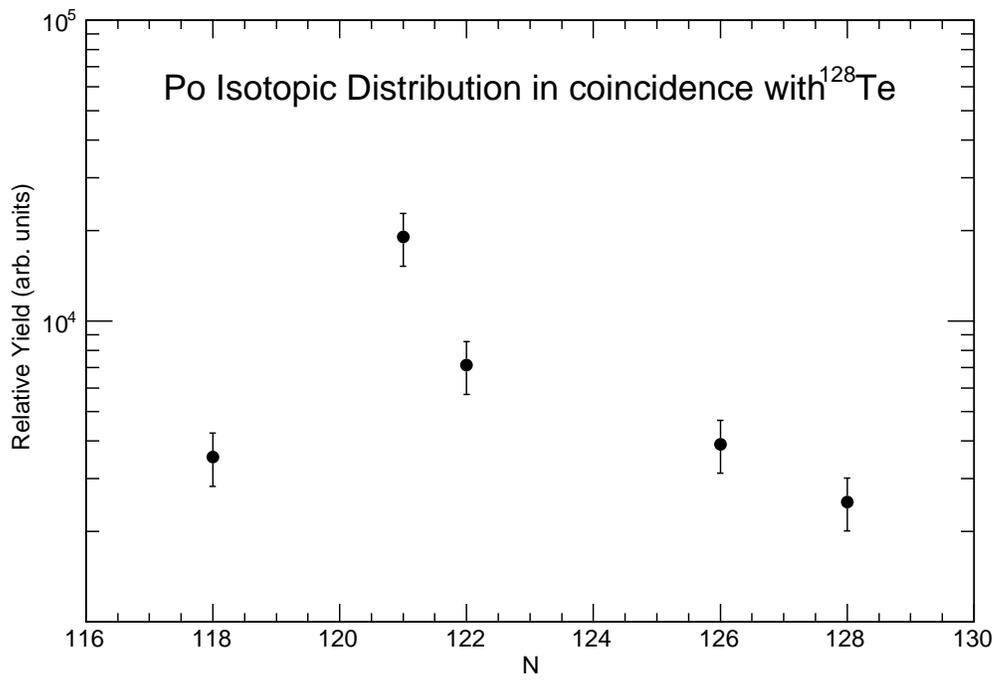}
\end{center}
\caption{\label{fig12}Distribution of TLFs produced when the PLF is $^{128}$Te.}
\end{figure}

% tables should appear as floats within the text
% The ruledtabular enviroment adds doubled rules to table and sets a
% reasonable default table settings.
% Use the table* environment to get a full-width table in two-column
% Add \usepackage{longtable} and the longtable (or longtable*}
% environment for nicely formatted long tables. Or use the the [H]
% placement option to break a long table (with less control than 
% \begin{table}%[H] add [H] placement to break table across pages
% \begin{ruledtabular}
% Lines of table here ending with \\
% \end{tabular}
% \end{ruledtabular}
% \end{table}

% Surround table environment with turnpage environment for landscape
% table
% \begin{table}
% \begin{ruledtabular}
% \end{tabular}
% \end{ruledtabular}
% \end{table}

%\appendix
%\section{}

%\begin{acknowledgments}
%\end{acknowledgments}


\end{document}